%% file: main.tex
\documentclass[9pt,conference]{IEEEtran}

\usepackage{multirow}
\usepackage{comment}
\usepackage{soul,xcolor,pifont}
\usepackage{colortbl}
\usepackage{array}
\usepackage{booktabs}
\usepackage{lipsum}
\usepackage{courier}
\usepackage{arydshln}
\usepackage{url}
\usepackage{hyperref}
\usepackage{amsmath,bm}
\usepackage{cleveref}
\usepackage{graphicx}
\usepackage[linesnumbered,ruled,vlined]{algorithm2e}
\usepackage{algpseudocode}
\usepackage{listings}
\usepackage{tikz}
\usepackage{mdframed}
\usepackage{boxedminipage}
\usepackage{setspace}
\usepackage{microtype}
\usepackage{courier}
\usepackage{etoolbox}

\usepackage{enumitem}
\usepackage{txfonts}

\def\BibTeX{{\rm B\kern-.05em{\sc i\kern-.025em b}\kern-.08em
	    T\kern-.1667em\lower.7ex\hbox{E}\kern-.125emX}}

\renewcommand{\IEEEbibitemsep}{0pt plus 0.5pt}
\makeatletter
\IEEEtriggercmd{\reset@font\normalfont\fontsize{7.7pt}{8.40pt}\selectfont}
\patchcmd{\@makecaption}
{\scshape}
{}
{}
{}
\makeatletter
\patchcmd{\@makecaption}
{\\}
{.\ }
{}
{}
\patchcmd{\@makecaption}
{\centering}
{}
{}
{}
\makeatother
\IEEEtriggeratref{1}
\def\tablename{Table}

\hypersetup{colorlinks=true,linkcolor=blue,citecolor=blue,urlcolor=magenta}

\setstcolor{red}

\definecolor{dkgreen}{rgb}{0,0.6,0}
\definecolor{mauve}{rgb}{0.58,0,0.82}
\definecolor{background}{rgb}{0.95,0.95,0.92}
\lstdefinestyle{mystyle}{
    basicstyle=\scriptsize\ttfamily,
    tabsize=1,
    captionpos=b, 
    frame=single,
    breaklines=true,
    language=c,
    keywordstyle=\color{blue},
    commentstyle=\color{dkgreen},
    stringstyle=\color{mauve},
    numbers=left,
    numberstyle=\tiny,
    numbersep=5pt,
}
\SetCommentSty{mycommfont}

\newcolumntype{M}[1]{>{\centering\arraybackslash}m{#1}}

\newcommand*\colourcheck[1]{
    \expandafter\newcommand\csname #1check\endcsname{\normalsize \textcolor{#1}{\ding{52}}}
}
\colourcheck{blue}
\colourcheck{green}
\colourcheck{red}
\colourcheck{black}

\newcommand*\colourcancel[1]{
    \expandafter\newcommand\csname #1cancel\endcsname{\normalsize \textcolor{#1}{\ding{56}}}
}
\colourcancel{blue}
\colourcancel{green}
\colourcancel{red}
\colourcancel{black}

\tikzset{
  vertical align/.style={
    line width=1.0,
    baseline=-.5*(height("$+$")-depth("$+$"))
  }
}

\begin{document}


\title{AVX Timing Side-Channel Attacks against Address Space Layout Randomization}
\author{\IEEEauthorblockN{Hyunwoo Choi}
\IEEEauthorblockA{\textit{The Affiliated Institute of ETRI} \\
zemisolsol@nsr.re.kr}
\and
\IEEEauthorblockN{Suryeon Kim}
\IEEEauthorblockA{\textit{KAIST} \\
c16192@kaist.ac.kr}
\and
\IEEEauthorblockN{Seungwon Shin}
\IEEEauthorblockA{\textit{KAIST} \\
claude@kaist.ac.kr}
}



\maketitle
\IEEEpubidadjcol

\renewcommand{\sectionautorefname}{Section}
\renewcommand{\algorithmautorefname}{Algorithm}
\newcommand{\aref}[1]{\hyperref[#1]{Appendix~\ref*{#1}}}
\crefformat{section}{\S#2#1#3}
\crefformat{subsection}{\S#2#1#3}
\crefformat{subsubsection}{\S#2#1#3}

\begin{abstract}
\input{abs}
\end{abstract}


\input{intro}
\input{back}
\input{anal}
\input{kaslr}
\input{disc}
\input{related}
\input{concl}

\section*{Acknowledgment}
We thank the anonymous DAC 2023 reviewers for their insightful feedback.
We would also like to thank our colleagues at the National Security Research Institute, particularly Dr. Jaeseo Lee for his assistance in publishing our paper.

\bibliographystyle{IEEEtranS}
\bibliography{refs}

\end{document}

%% file: abs.tex
Modern x86 processors support an AVX instruction set to boost performance.
However, this extension may cause security issues.
We discovered that there are vulnerable properties in implementing masked load/store instructions.
Based on this, we present a novel AVX timing side-channel attack that can defeat address space layout randomization.
We demonstrate the significance of our attack by showing User and Kernel ASLR breaks on the recent Intel and AMD processors in various environments, including cloud computing systems, an SGX enclave (a fine-grained ASLR break), and major operating systems.
We further demonstrate that our attack can be used to infer user behavior, such as Bluetooth events and mouse movements.
We highlight that stronger isolation or more fine-grained randomization should be adopted to successfully mitigate our presented attacks.

%% file: intro.tex
\section{Introduction}
\label{sec:intro}

Modern x86 processors support a Single Instruction Multiple Data (SIMD) instruction set that compilers or programmers can use to boost performance~\cite{intel2021optz,mediainst2021amd}.
However, this instruction set may also cause security issues.
In particular, we discovered that in the Advanced Vector Extensions (AVX), there are vulnerable properties in the implementation of the masked load/store instructions.
First, these instructions can suppress exceptions caused by invalid or inaccessible memory access.
Second, the execution time of these instructions leaks the current state of the page mappings and permissions, as well as TLB states.

In this paper, based on these vulnerable properties, we introduce a novel AVX timing side-channel attack that can defeat address space layout randomization (ASLR).
We demonstrate the significance of our attack by showing User and Kernel ASLR breaks on both recent Intel and AMD CPUs.
Specifically, we show that our attack reliably retrieves the base address of the Linux kernel text in 0.28 $ms$ with a near-zero error rate.
For kernel modules, based on a unique size, our attack can identify the currently loaded modules in 2.62 $ms$ with 99.84\% accuracy.
The attack works even on a kernel page table isolation (KPTI)-enabled kernel.
This also shows that our attack can be used to infer user behavior by monitoring kernel activities such as Bluetooth events and mouse movements.
Furthermore, we demonstrate the applicability of our attack by showing KASLR breaks in various environments, including cloud computing systems, a Software Guard Extensions (SGX) enclave, and other major operating systems (OSes).
In particular, we show how our attack can be used to mount a fine-grained ASLR break inside an SGX enclave.

Since our attack only requires AVX instructions, it makes it much more practical compared to known microarchitectural attacks that depend on noise filtering~\cite{hund2013practical}, hardware transactional memory (Intel TSX)~\cite{jang2016breaking}, cache eviction~\cite{gruss2016prefetch,lipp2022amd}, knowledge of the branch target buffer (BTB) hash function~\cite{evtyushkin2016jump}, the TLB addressing~\cite{koschel2020tagbleed}, cache status monitoring~\cite{canella2019fallout,canella2020kaslr,weber2021osiris}, or the energy reporting interface (e.g., RAPL)~\cite{lipp2021platypus,lipp2022amd}.

In summary, our main contributions are as follows:
\setlist[itemize]{leftmargin=3mm}
\begin{itemize}
	\setlength\itemsep{0.1em}
  \item We conduct an in-depth analysis of the AVX masked operations and discover vulnerable properties in the implementation of masked load and store instructions.
	\item We present a novel AVX timing side-channel that can defeat address space layout randomization. We demonstrate User and Kernel ASLR breaks on the recent Intel and AMD CPUs.
	\item We show that our attack can detect user behavior and is feasible in cloud computing systems (Amazon EC2, Google GCE, and Microsoft Azure), an SGX enclave (a fine-grained ASLR break), and major OSes (Linux and Windows).
  \item We responsibly disclosed our findings to Intel on April 20, 2022, and AMD on July 3, 2022. Intel acknowledged them on May 10, 2022, and AMD on July 5, 2022.\footnote{The proof-of-concept code of our attack is available at \url{https://github.com/zemisolsol/kaslrAVX}.}
\end{itemize}

%% file: back.tex
\section{Background}
\label{sec:back}

\subsection{Address translation and translation caches}

\noindent\textbf{Address translation.}
Virtual addresses are translated into physical addresses through multi-level page tables.
On Intel x86-64 processors, page tables are comprised of four levels of paging structures: page map level 4 (PML4), page directory pointer table (PDPT), page directory (PD), and page table (PT).
Each page table defines the mappings from a virtual address to a physical address, and address translation is performed by indexing certain parts of the virtual address.
The virtual address space is divided into user space and kernel space, which serves to provide memory protection.
To this end, the page tables contain permission-related information, such as a readable/writable page and a user/kernel accessible page.

\noindent\textbf{Translation Lookaside Buffer.}
TLB is a special cache that contains the most recently used page table entries (PTEs).
Given a virtual address, the processor examines the TLB.
If a PTE is present (called a \textit{TLB hit}), the corresponding page frame number (PFN) is retrieved and the physical address is formed.
If the PFN is not found (a \textit{TLB miss}), the processor starts to walk the page table hierarchy (called a \textit{page table walk}) to look for the corresponding PFN.
On page table walks, a memory management unit (MMU) accesses each page table to find the translation for the virtual address.
Once the translation is found, meaning that the page is mapped and no page fault (\textit{\#PF}) occurs, the TLB is updated to include the new page entry.
Otherwise, \textit{\#PF} is issued, and the OS handles the \textit{\#PF}.

\noindent\textbf{Page-translation caches.}
The page table can still be cached just like any other read from normal memory.
However, frequently accessing the PTEs on every page walk will incur a penalty of several tens of cycles per TLB miss, even if all entries are present in the data cache (e.g., L2).
For this reason, modern processors may have page-translation caches (Intel refers to these as \textit{paging-structure caches}~\cite{intelsdmvol3a}) to further improve the performance of the TLB miss~\cite{barr2010translation}.

\subsection{KASLR and Kernel Page-Table Isolation}

\noindent\textbf{Kernel Address Space Layout Randomization.}
KASLR is a technique used to randomize the base address of a kernel image and the position of kernel modules at a boot or driver load time.
Once KASLR is enabled, it defeats code reuse attacks such as return-oriented programming (ROP)~\cite{shacham2007geometry}, which rely on  knowledge of the absolute address of instructions.
In an x86-64 Linux kernel, the kernel image is aligned to a 2 MiB boundary and mapped between \texttt{0xffffffff}\textbf{\texttt{800}}\texttt{00000}-\texttt{0xffffffff}\textbf{\texttt{c00}}\texttt{00000} with a maximum size of 1 GiB (i.e., 512 possible offsets).
Although the randomization's entropy is only 9 bits, brute force attacks against the kernel are virtually not feasible due to the high rate of a kernel panic.

\noindent\textbf{Kernel Page-Table Isolation.}
KPTI was first introduced to defend against attacks on KASLR~\cite{gruss2017kaslr}.
With KPTI, kernel space is isolated from user space; thus, it undermines attacks that are based on the status of page mappings.
Major OSes have adopted this isolation technique as mitigation for a Meltdown attack (Linux PTI~\cite{kpti2017lwn}, Microsoft Kernel Virtual Address Shadow (KVAS)~\cite{windowskvashadow}, and Apple Double Map~\cite{appledoublemap}).

\subsection{Advanced Vector Extensions}

AVX is a SIMD instruction set supported by Intel and AMD processors.
With AVX, arithmetic and data transfer operations can be processed simultaneously.
Although modern compilers such as GNU GCC and Intel C++ provide automatic vectorization options (e.g., \texttt{/Qvec} in Intel C++ compiler~\cite{deilmann2012guide}), an advanced user can obtain better performance with AVX programming.
As one of the optimizations in AVX, the masked load/store operations (\texttt{VMASKMOV} and \texttt{VPMASKMOV}) are used to conditionally move packed data elements to/from memory, depending on the mask bits associated with each data element.
In this paper, we exploit vulnerable properties of the masked operations to mount timing side-channel attacks.

%% file: anal.tex
\begin{figure}[t]
  \centering
  \includegraphics[scale=0.41]{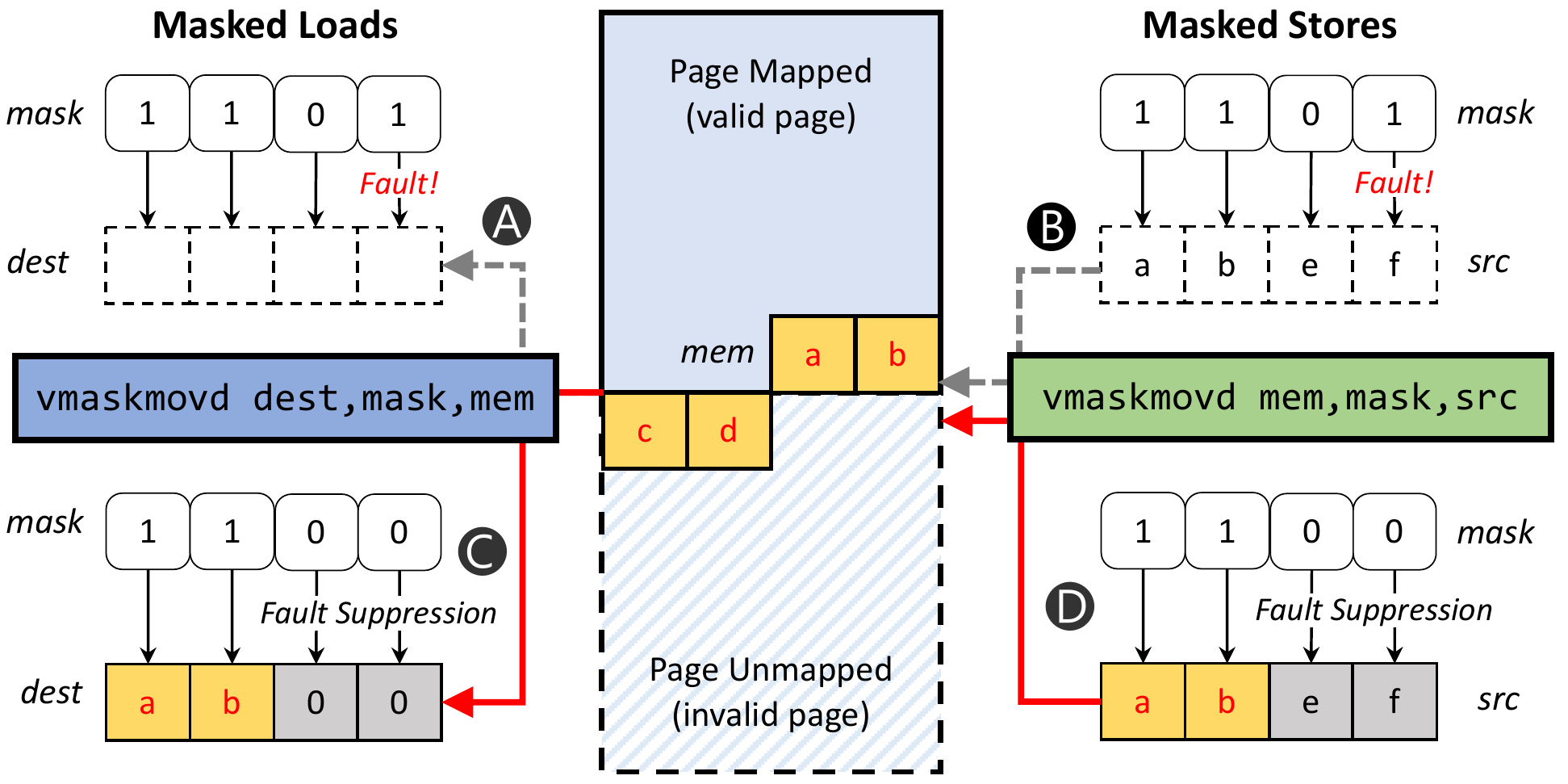}
  \caption{A fault suppression of the AVX masked load/store. 
    On unmapped pages, no faults are issued if the mask bits are all set to ``\textit{zero}" (C and D).}
  \label{fig:faultsuppression} 
\end{figure}

\section{In-depth analysis of the AVX timing side-channel}
\label{sec:anal}

This section first analyzes the vulnerable properties of the AVX masked load/store instruction and then presents three attack primitives.

\subsection{Fault-resistance}

Intel optimization manual~\cite{intel2021optz} describes a fault-resistance property of the AVX masked operations.
To verify that a masked load/store instruction does indeed suppress the exception, we examined memory access on an Intel i7-1065G7 (Ice Lake) CPU (\autoref{fig:faultsuppression}).
We prepared two adjacent pages using the \texttt{mmap}/\texttt{munmap} syscalls; the upper page is mapped (a valid page), while the lower page is unmapped (an invalid page).
We first executed the masked load/store across the page boundary, where only one element on the low page is masked (\textbf{A} and \textbf{B}).
We then examined the case where all the elements on the lower page are masked out (\textbf{C} and \textbf{D}).
The experiment revealed that when accessing an unmapped page, no faults occur if the corresponding mask bits are all set to ``\textit{zero}".
We further tested the masked load/store on kernel memory to determine whether it applies to inaccessible pages.
As a result, no faults were observed.
Thus, we prove that when executing the masked load/store instruction, masking out does indeed suppress the exceptions even if the page being accessed is invalid or inaccessible.

\begin{mdframed}
\hspace{-0.25cm}$\varmathbb{P}1$: \textit{The AVX masked operations can suppress the exceptions caused by invalid or inaccessible memory accesses.}
\end{mdframed}

\subsection{Timing differences}
\label{subsec:property}


\begin{figure}[t]
  \centering
  \includegraphics[scale=0.365]{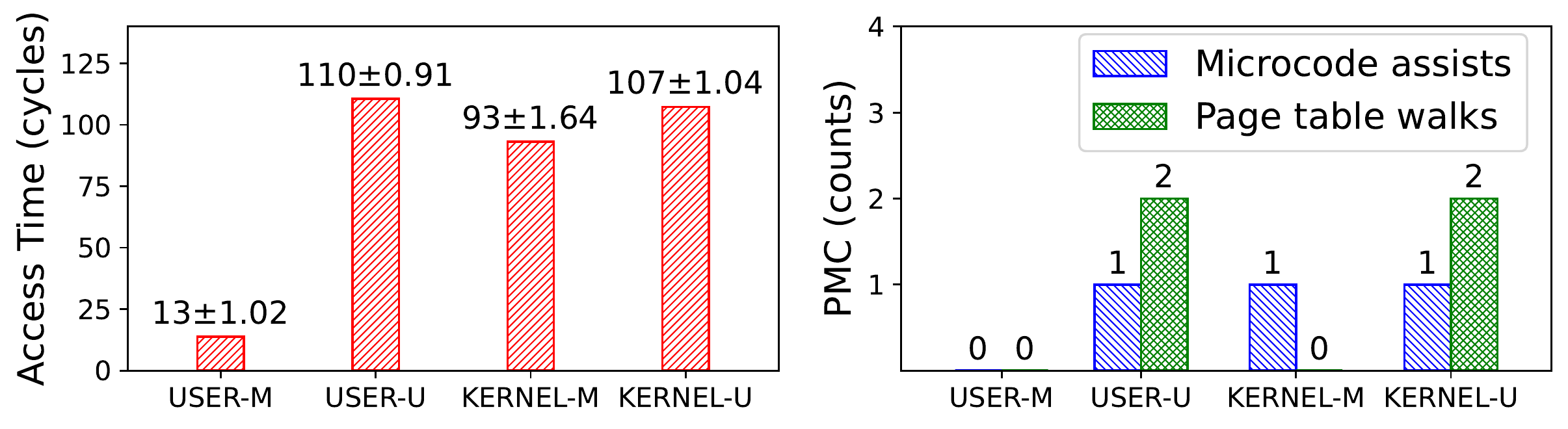}
  \caption{Execution times of the masked load instructions for different types of pages on an Intel i7-1065G7 (Ice Lake) (on the left side). 
    It also shows the number of corresponding performance counters (on the right side).}
  \label{fig:pmc}
\end{figure}

\noindent\textbf{Page-Table Level.}
The masked load/store instructions trigger a \textit{microcode assist} when the address being accessed is invalid or inaccessible~\cite{intel2021optz}. 
The microcode assist may take additional cycles because it may need to determine whether the elements have the corresponding mask bits set.
To verify this, we measured the execution time of the masked load for different types of pages: {USER-M} (a page in the user space with Present-bit:1 and User/Supervisor-bit:1), {USER-U} (P:0), {KERNEL-M} (a page in the kernel space with P:1 and U/S:0), and {KERNEL-U} (P:0).
We also measured their corresponding microcode assist events (\texttt{ASSISTS.ANY}) using a performance counter monitor.
\autoref{fig:pmc} depicts the measurement results.
On the {USER-M} page, without issuing a microcode assist, the mean value of an access time is 13 cycles.
In contrast, on other pages, the access time is significantly increased due to the microcode assist.

In particular, we observed that the {KERNEL-M} has a shorter access time ($<$ 14 cycles) than the {KERNEL-U}.
Since the address translation of the unmapped page (P:0) may not be stored in the TLB, we can speculate that the timing difference between the two pages is due to the page table walks.
To prove this, when accessing the kernel address, we measured the number of completed page table walks (\texttt{DTLB\_LOAD\_MISSES.WALK\_COMPLETED}).
As we expected, the page table walks were triggered twice in the KERNEL-U but not in KERNEL-M (right in~\autoref{fig:pmc}).
Therefore, we prove that the execution time of the masked load/store on kernel-mapped pages is faster than on unmapped pages.

\begin{mdframed}
\hspace{-0.25cm}\textbf{$\varmathbb{P}2$}: \textit{The masked operations can distinguish between mapped and unmapped pages by measuring execution time.}
\end{mdframed}

The execution time of a page table walk varies depending on the level of the page table where the walk terminates~\cite{gruss2016prefetch,lipp2022amd}.
To verify that the masked load/store can also leak information about a page table's level, we measured its execution time on different levels of page tables on the Intel i-9900 (Coffee Lake).
In Ubuntu 20.04.4 (kernel 5.13.0-30), we executed the masked load instruction with four different kernel addresses that are mapped to PT, PDT, PDPT, and PML4T, respectively.
Since the translation of a valid address is cached in the TLB, we flushed the TLB (using an \texttt{INVLPG} instruction in LKM) before the measurement to trigger page table walks.
As a result, we observed that the execution time increases linearly from the lowest level (PDT) to the highest level (PML4T) except for PT.
Note, as Intel's paging-structure caches do not contain PT, walking page tables takes longer when translating a virtual address mapped on a 4 KiB page (PT) compared to huge pages.

\begin{mdframed}
\hspace{-0.25cm}\textbf{$\varmathbb{P}3$}: \textit{The masked operations can leak information about the level of the page table where the walk terminates.}
\end{mdframed}

\noindent\textbf{TLB state.}
The execution time of the masked load/store instruction differs depending on the TLB state.
If a page table entry is present in the TLB during the address translation (i.e., a TLB hit), it takes less time than a TLB miss (see~\cref{sec:back}).
To verify this, we tested memory access on the kernel-mapped page to determine whether the masked operations can distinguish between a TLB hit and miss.
On the Intel processor with the {KERNEL-M} page, we executed the masked load instruction twice in a row (first for a TLB miss and then for a TLB hit) and measured each execution time.
Before the first access, we evicted the TLB entries~\cite{gras2018translation} to ensure that the first execution issues a TLB miss.
We repeated this test 1000 times on an Intel i9-9900 (Ubuntu 20.04.1 with kernel 5.11.0-27).
The experimental results show that the first execution takes an average of 381 cycles, while the second execution takes 147 cycles, on average.
As a result, we confirm that the masked operations can be used to identify the current TLB state.

\begin{mdframed}
\hspace{-0.25cm}\textbf{$\varmathbb{P}4$}: \textit{The masked operations can identify TLB states.}
\end{mdframed}

\begin{figure}[t]
  \centering
  \includegraphics[scale=0.35]{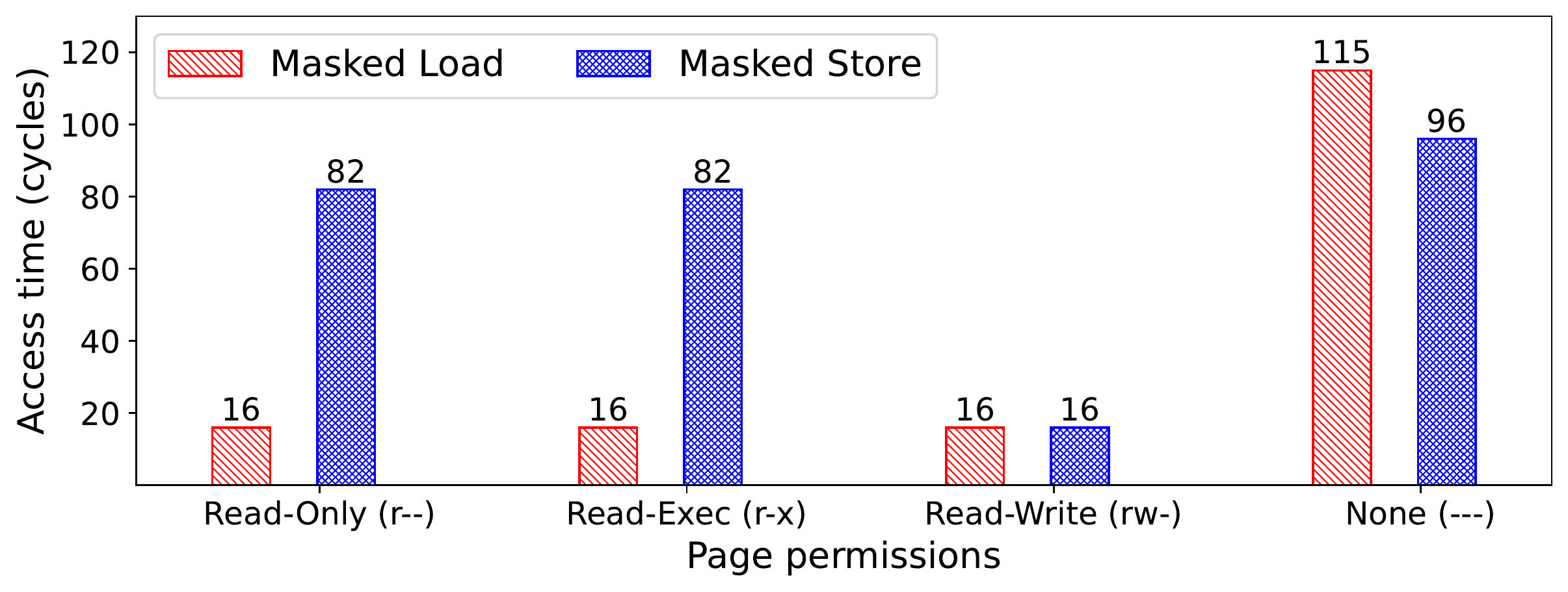}
  \caption{A comparison of execution times according to page permissions. 
		A masked load can distinguish two types of pages (\texttt{r--/r-x} and \texttt{---}), whereas a mask store can distinguish three types of pages (\texttt{r--/r-x}, \texttt{rw-}, and \texttt{---}).}
  \label{fig:permissions}
\end{figure}

\noindent\textbf{Page permission.}
The execution time of the masked load/store instruction is affected by page permissions being accessed.
To evaluate this, we mapped four pages with different permissions (\texttt{read-only}, \texttt{read-exec}, \texttt{read-write}, and \texttt{none}) in the user space and measured each execution time of the masked load/store instructions. 
As a result, we observed that the execution time of the masked load differs only in the \texttt{none} page permission (\autoref{fig:permissions}).
However, in the case of the masked store, we observed that write permission (\texttt{read-write}) affects the execution time.
If a page does not have the \texttt{write} permission, the masked store triggers a microcode assist, which takes additional cycles.
Thus, in the execution of the masked store, the timing differences between read (\texttt{read-only} and \texttt{read-exec})  and write (\texttt{read-write}) permissions are clearly visible.

\begin{mdframed}
\hspace{-0.25cm}\textbf{$\varmathbb{P}5$}: \textit{The masked operations can identify page permissions.}
\end{mdframed}

\noindent\textbf{Load and store.}
The masked load and store have most of the same properties discussed above, except for the execution time.
On an Intel i7-1065G7, we executed the masked load and store instructions on the KERNEL-M page and measured each execution time.
The masked load takes an average of 92 cycles, while the execution time of the masked store is 76 cycles.
The results reveal that the masked store takes roughly 16-18 cycles less time to execute than the masked load.

\begin{mdframed}
\hspace{-0.25cm}\textbf{$\varmathbb{P}6$}: \textit{The masked store executes faster than the masked load.}
\end{mdframed}

\subsection{Attack primitives}


\noindent\textbf{Page-table attack.}
The page-table attack can distinguish between the present (valid or mapped) and non-present (invalid or unmapped) pages ($\mathbb{P}2$) or directly leak the page-table level of the present pages at which the page-table walk terminates ($\mathbb{P}3$). 
In this paper, we show how to reliably break KASLR on both Intel and AMD CPUs based on the page-table attack.
\vspace{0.5em}

\noindent\textbf{TLB attack.}
The TLB attack can identify current TLB states, i.e., a TLB hit or miss ($\mathbb{P}4$).
We use this attack primitive to detect user behavior by measuring the execution time of the masked operation on the kernel modules.
Note, we use this attack primitive in combination with a TLB eviction to reduce noise.
We further use the TLB attack to bypass FLARE~\cite{canella2020kaslr}, a state-of-the-art defense against currently known KASLR breaks.
\vspace{0.5em}

\noindent\textbf{Permission attack.}
The permission attack can identify the current page permissions ($\mathbb{P}5$).
With this attack primitive, we can identify whether the page is readable or writable.
Since the Linux kernel adopts strict kernel memory permissions where any area of the kernel with executable memory must not be writable~\cite{strictkernelmemory}, in this paper, we use this attack primitive to implement a fine-grained ASLR break in the user address space (even inside an SGX enclave).

Note that all of our attack primitives suppress page faults caused by invalid or inaccessible memory addresses ($\mathbb{P}1$).

%% file: kaslr.tex
\section{AVX timing side-channel attacks}
\label{sec:kaslr}

This section shows how the AVX timing side-channel can be used to defeat User and Kernel ASLR on recent Intel and AMD CPUs.

\subsection{Threat model}
\label{sec:threat}

We assume an unprivileged attacker that executes arbitrary instructions on the User/Kernel ASLR-enabled local machine.
The attacker's goal is to know the addresses of the codes to attempt code reuse attacks with the knowledge of the CPU model and kernel functions' constant offsets.
We assume that there are no software-based memory leak vulnerabilities.
For hardware, we assume that the processor supports AVX or AVX2 and is protected by mitigations against the existing side-channel attacks~\cite{intel2018mitigations}.
Since AVX was introduced in 2011 on Intel/AMD CPUs, it is reasonable to assume that the vast majority of systems support AVX by default.

\subsection{Derandomizing the kernel base address}
\label{subsec:kaslrbase}

\begin{figure}[t]
  \centering
  \includegraphics[scale=0.395]{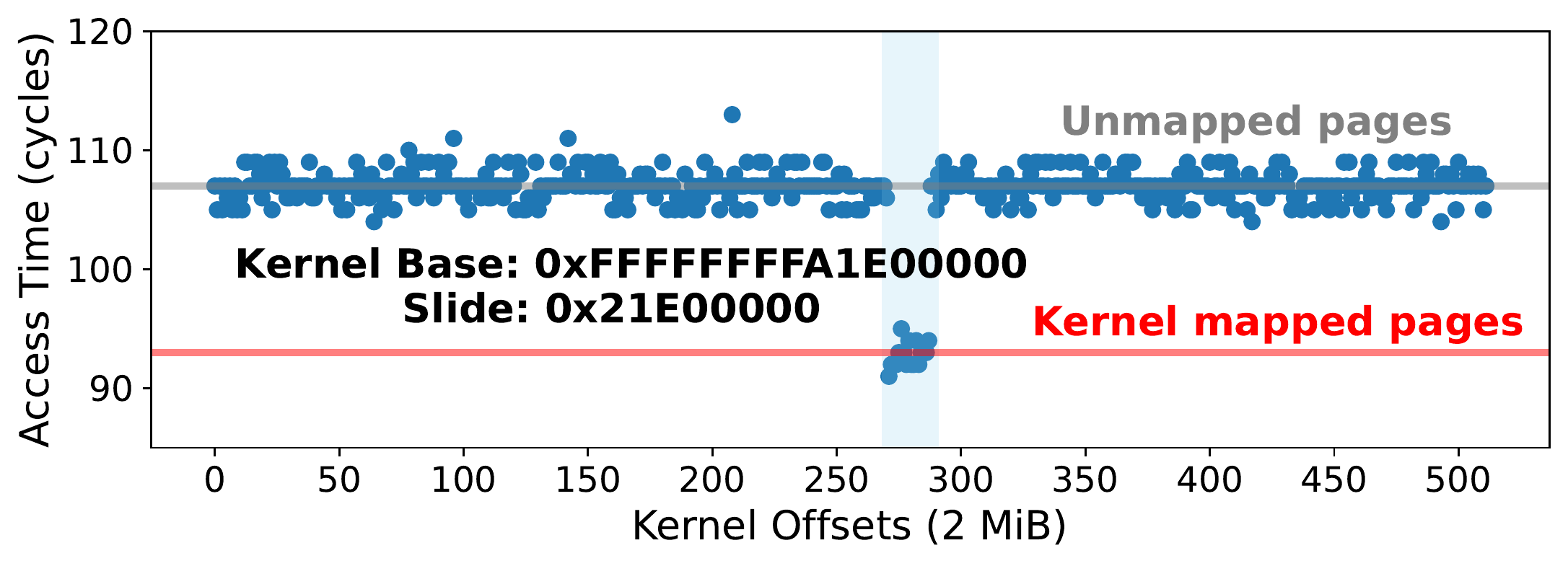}
  \caption{Measurement results of the probing kernel address range on Linux with an Intel i5-12400F (Alder Lake). 
    The lower plots show the execution times (average: 93 cycles) of kernel-mapped pages.}
  \label{fig:baseaddress}
\end{figure}

We demonstrate KASLR break on Ubuntu 20.04.3 (kernel 5.11.0-27) with a Meltdown-resistant Intel i5-12400F (Alder Lake).
With KASLR, the base address of the Linux kernel is located in 512 possible offsets (see~\cref{sec:back}).
In our attack, we measure the execution time of the masked load instruction on each possible offset.
First, we determine a threshold value to distinguish between mapped and unmapped pages.
We found that the execution time of the masked store on the user-mapped page with no dirty bit set (D:1) is the same as the execution time on the kernel-mapped page.
Thus, we use the average execution time of the masked store on the {USER-M} page as our threshold value.
Next, we execute the masked load instruction twice for each of the 512 candidate addresses and measure the execution time of the second execution ($\varmathbb{P}2$).
As a result, we clearly identified kernel-mapped addresses in the user space (\autoref{fig:baseaddress}).
Note, our attack suppresses page faults caused by accessing inaccessible or invalid kernel addresses ($\varmathbb{P}1$).

The measurement results are shown in~\autoref{fig:baseaddress}.
From 512 plots, we can clearly distinguish between the execution times for kernel-mapped and unmapped pages.
The kernel-mapped pages have a mean execution time of 93 cycles, while the unmapped pages have 107 cycles.
Since the lower plots start at offset 271, we can identify the base address of the kernel (\texttt{0xffffffffa1e00000}).
To verify the result, we rebooted Linux 10 times and checked whether the identified base address is correct by confirming a \texttt{/proc/kallsyms} file.
In each attempt, we always found the correct base address of the kernel without any false positives.
The average runtime of probing the kernel address range is 0.67 $\mu$$s$, while the total average runtime is 0.28 $ms$ (\autoref{tbl:evalbase}).
The accuracy of the attack is 99.6$\%$, on average (\textit{n = 10000}).

We further performed the attack on an AMD Ryzen 5 5600X (Zen 3).
On AMD, we observed that accessing kernel addresses always triggers page table walks regardless of page mappings.
Thus, to break KASLR on AMD, we used a page-table attack ($\varmathbb{P}3$) and exploited the fact that Linux's kernel-mapped area contains 4-KiB pages~\cite{lipp2022amd}.
In our experiment, our attack reliably identified five 4-KiB pages within the kernel address range.
We achieved an average runtime of 2.9 ms with 99.48\% accuracy (\textit{n = 10000}).
We leave further evaluation, such as kernel base and module detection on various AMD CPUs, for future work.

\begin{table}[t]
\renewcommand{\arraystretch}{1.1}
\caption{An average runtime and accuracy for derandomizing kernel base and module addresses. The \textit{Probing} runtime is the time it takes to execute the masked operations only.}
\label{tbl:evalbase}
\centering
\scalebox{0.65}{
\begin{tabular}{M{3.7cm}M{1.4cm}M{1.5cm}M{1.5cm}M{1.4cm}}
\toprule
\multicolumn{1}{l}{\multirow{2.5}{*}{\textbf{CPUs (setting, launch date)}}}  & \multicolumn{1}{c}{\multirow{2.5}{*}{\textbf{Targets}}} & \multicolumn{2}{@{}c@{}}{\textbf{Runtime (\textit{n = 10000})}}  &  \multicolumn{1}{c}{\multirow{2.5}{*}{\textbf{Accuracy}}} \\ \cmidrule(rl){3-4} 
\multicolumn{1}{l}{} & \multicolumn{1}{l}{} & \multicolumn{1}{c}{\textbf{Probing}} & \multicolumn{1}{c}{\textbf{Total}}  & \multicolumn{1}{l}{}\\
\midrule \midrule
\multicolumn{1}{l}{\multirow{2.5}{*}{Intel Core i5-12400F (Desktop, Q1'22)}}        & Base    & 67 $\mu$$s$ & 0.28 $ms$ & 99.60 $\%$ \\
\cmidrule(rl){2-5} 
     & Modules & 2.43 $ms$ & 2.62 $ms$ & 99.84 $\%$ \\
\midrule
\multicolumn{1}{l}{\multirow{2.5}{*}{Intel Core i7-1065G7 (Mobile, Q3'19)}}  & Base    & 0.26 $ms$ & 0.57 $ms$ & 99.29 $\%$ \\
\cmidrule(rl){2-5} 
 & Modules & 8.42 $ms$ & 8.64 $ms$ & 99.72 $\%$ \\
\midrule
\multicolumn{1}{l}{AMD Ryzen 5 5600X (Desktop, Q2'20)} & Base    &  1.91 $ms$ &  2.90 $ms$ &  99.48 $\%$ \\
\bottomrule
\end{tabular}
}
\end{table}

\subsection{Detecting and identifying kernel modules}
\label{subsec:module}

In the x86-64 Linux, kernel modules (or drivers) are loaded between \texttt{0xffffffffc}\textbf{\texttt{0000}}\texttt{000}-\texttt{0xffff}\texttt{ffffc}\textbf{\texttt{40}\texttt{00}}\texttt{000}, with a 4 KiB alignment. 
Thus, by probing the address range with 4-KiB offsets (16384 possible addresses), our attack can identify the addresses of the currently loaded modules.
For the attack, we first extract all mapped pages in the address range of the kernel modules by measuring timing differences (\textbf{$\varmathbb{P}2$}).
Then, as in prior work~\cite{canella2019fallout}, we distinguish where a module begins and ends by taking advantage of the fact that the loaded kernel modules are separated by unmapped pages. 
As a result, we can identify all loaded modules and their size. 
Since Linux's \texttt{/proc/modules} file provides module information such as name, and size, we can classify modules by correlating the detected module size with the actual size.

\begin{figure}[t]
  \centering
  \includegraphics[scale=0.395]{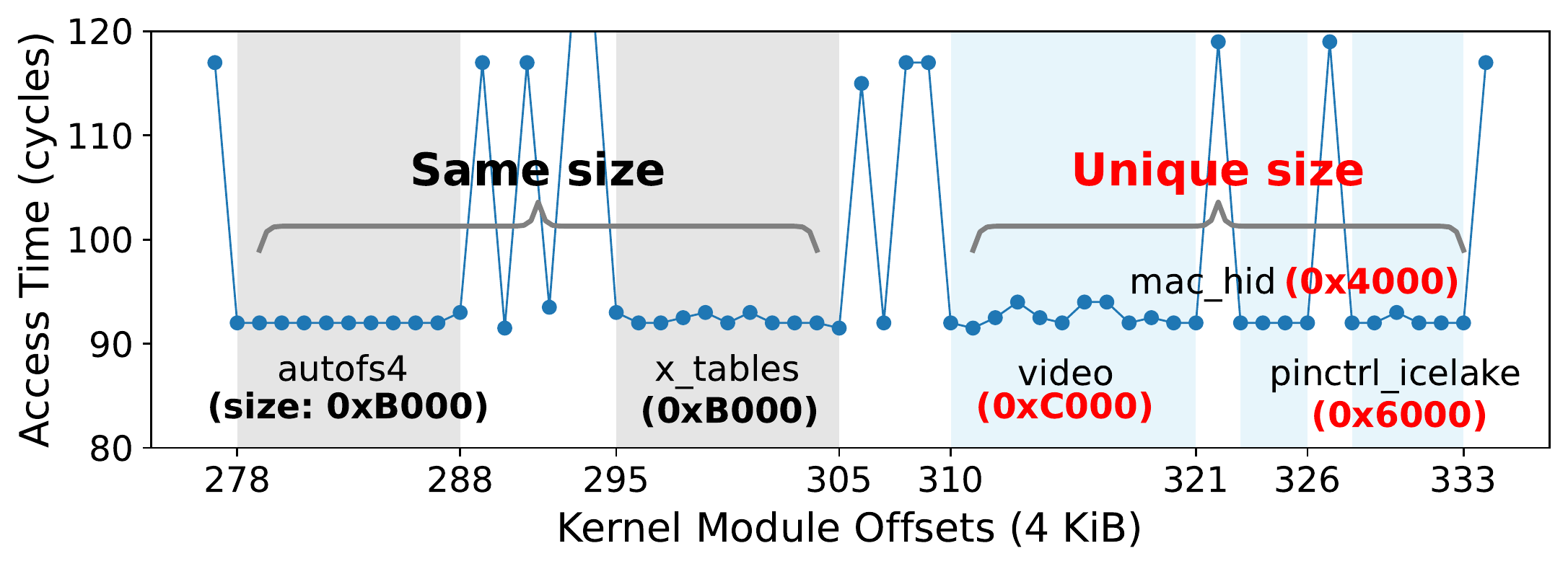}
  \caption{Identified kernel modules' offsets.
  The graph shows an example of five modules along with their names and sizes.
  \texttt{video}, \texttt{mac\_hid} and \texttt{pinctrl\_icelake} can be accurately identified by their unique sizes.}
  \label{fig:module}
\end{figure}

We evaluated our attack on Ubuntu 18.04.3 (kernel 5.4.0-81) with an Intel i7-1065G7 (Ice Lake), where the total number of loaded kernel modules is 125, of which 19 have a unique size.
\autoref{fig:module} depicts an example of the identified five kernel modules along with their names and sizes.
As our classification is based on the detected size, we cannot differentiate between \textit{autofs4} and \textit{x\_tables} that map with the same amount of pages.
However, we can identify \textit{video}, \textit{mac\_hid}, and \textit{pinctrl\_icelake}, which have unique sizes.
Our attack achieved 8.42-8.64 $ms$ of runtime and 99.72$\%$ accuracy, on average (\autoref{tbl:evalbase}).
Note, the performance results are greatly improved in a desktop setting (Intel i5-12400F).

\subsection{Breaking KASLR with KPTI enabled}
\label{subsec:kaslrkpti}

In a KPTI-enabled kernel, the kernel pages are not mapped in the user space.
However, to provide an entry point into the kernel space, the KPTI leaves a minimal set of kernel pages called \textit{KPTI trampoline} in the user space, which is used to switch between the user and the kernel space (e.g., \texttt{entry\_SYSCALL\_64} for a syscall entry point).
We can determine the base address of the kernel image using the addresses of the mapped KPTI trampoline pages since the randomization is performed by shifting the entire kernel image within a given range.

We evaluated our attack on a KPTI-enabled kernel (Ubuntu 20.04.3 with kernel 5.11.0-27).
We first fixed the kernel's base address at \texttt{0xffffffff81000000} via a boot parameter (\textit{nokaslr}), then we performed the page-table attack ($\mathbb{P}2$).
In repeated experiments, we observed that the fast execution time appears at \texttt{0xffff}\texttt{ffff}\texttt{81c00000}, which is the same result as the confirmed constant offset of the KPTI trampoline (\texttt{0xc00000}) beforehand.
As such, with the knowledge of the KPTI trampoline offset, our attack can still break KASLR even on the KPTI-enabled kernel.

\subsection{Inferring user behaviors}
\label{subsec:infer}

We demonstrate that our TLB attack can also be used to infer user behaviors, as has been shown in prior works~\cite{canella2019fallout,lipp2022amd}.
Specifically, we monitor two types of user activities: Bluetooth audio streaming and mouse movements.
To this end, we target two different kernel modules (\texttt{bluetooth} and \texttt{psmouse}) and keep track of their events by measuring TLB states ($\mathbb{P}4$).
When the module is accessed, the address translations will be cached in the TLB.
Thus, the execution time varies depending on whether the module is in use.

\begin{figure}[t]
  \centering
  \includegraphics[scale=0.39]{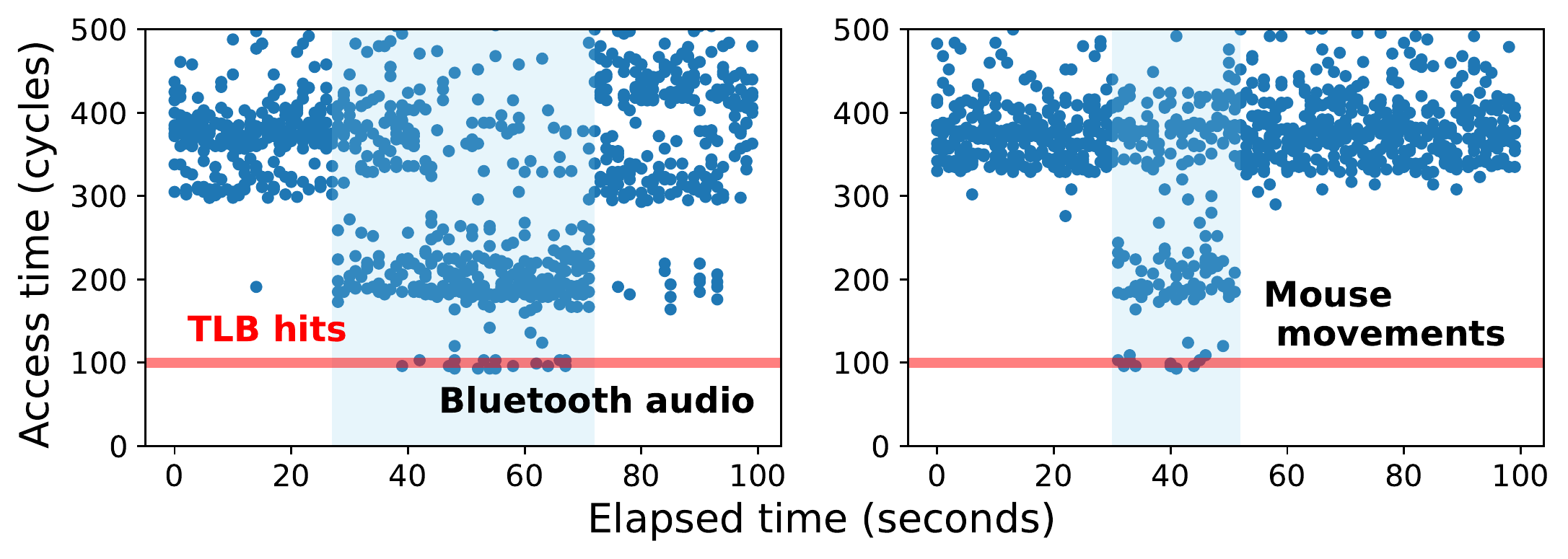}
  \caption{User behavior detection by measuring the TLB states of kernel modules (\texttt{bluetooth} and \texttt{psmouse}).}
  \label{fig:bluetooth}
\end{figure}

We performed the attack on Ubuntu 18.04.3 (kernel 5.4.0-81) with an Intel i7-1056G7.
In the experiment, a spy process repeats the TLB attack at 1 $sec.$ intervals and lasts for up to 100 $sec$.
\autoref{fig:bluetooth} shows the results obtained by the spy process measuring the masked load execution time on the first 10 pages of the kernel modules.
From the graphs, we can observe that the execution times are obviously shorter (blue area) when the kernel modules are accessed.
As such, attackers can utilize our attack to infer user behaviors to proceed with further attacks.
We believe that our attack will likely be extended not only to monitor other events (e.g., keystroke) but also to fingerprint applications or websites.

\subsection{Fine-grained ALSR break inside an SGX enclave}
\label{subsec:sgxaslr}

In x86-64 Linux, ASLR entropy for the process's address space is 28 bits.
For example, the process's code text is located within \texttt{0x55}\textbf{\texttt{XXXXXXX}}\texttt{000}, and the libraries are loaded within \texttt{0x7f}\textbf{\texttt{XXXXXXX}}\texttt{000}.
To explore such address spaces, we linearly probe the entire virtual address range with a 4-KiB alignment by measuring the execution time of the masked load/store instruction ($\varmathbb{P}2$).
On Ubuntu 18.04.3 (kernel 5.4.0-81) with an Intel i7-1065G7, our attack successfully identifies the base address of the process's code section inside an SGX enclave.
In our unoptimized proof-of-concept implementation, with SGX2 that supports a high-precision timer (\texttt{RDTSC} and \texttt{RDTSCP}), our attack takes on average 51 $sec.$ (masked load) and 44 $sec.$ (masked store).

\begin{figure}[t]
  \centering
  \includegraphics[scale=0.31]{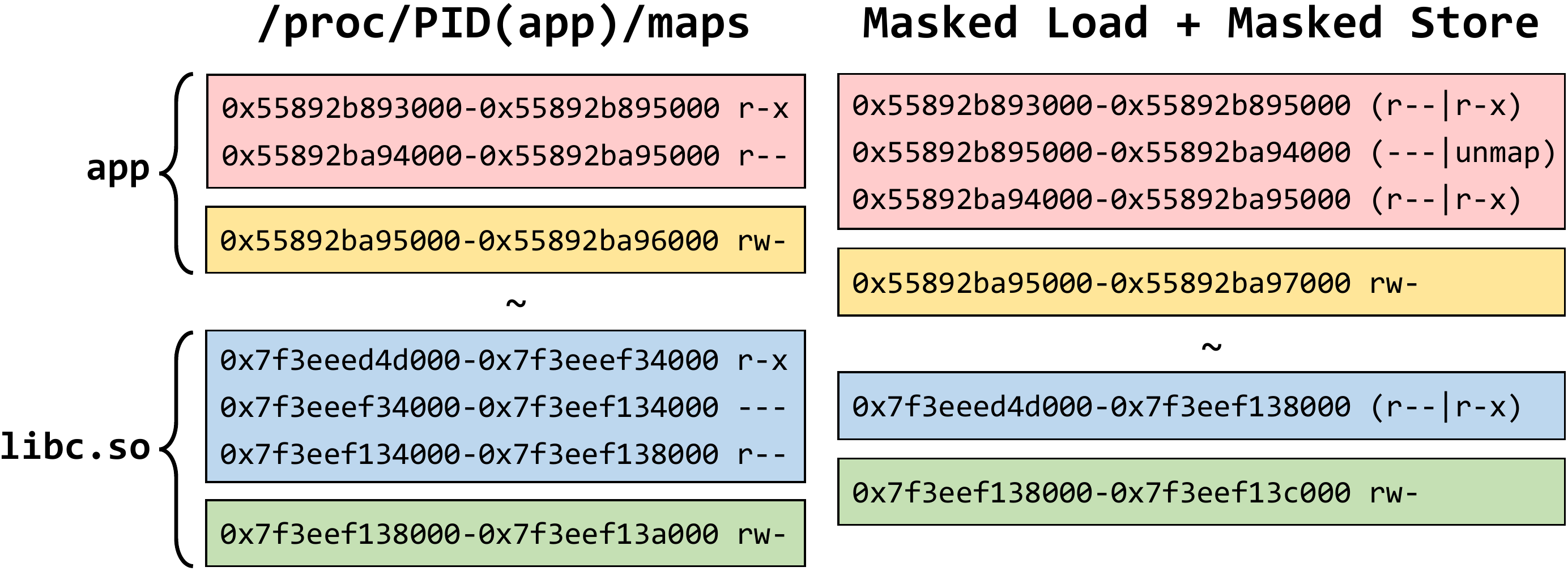}
  \caption{The process's identified mapped memory regions and their access permissions.
  The left is the output of the \texttt{maps} file, and the right is the result confirmed by our attack.
  For simplicity, we depict a \texttt{libc.so} library only.}
  \label{fig:aslr}
\end{figure}

To identify loaded libraries, we use a fine-grained methodology based on \textbf{$\varmathbb{P}5$}.
On Ubuntu 18.04.3, we observed that the loaded libraries (e.g., \texttt{libc.so}) consist of consecutive sections and the sections' permissions are in the order of \texttt{r-x}, \texttt{-}\texttt{-}\texttt{-}, \texttt{r-}\texttt{-}, and \texttt{rw-}.
With this, we used sections' sizes as signatures for detecting libraries.
To reduce noise, we probed the address space twice by combining the masked load and store.
We first probed the address space using the masked load and filtered out the \texttt{none} pages.
We then probed again using the masked store to identify the \texttt{read-write} pages.
\autoref{fig:aslr} shows the results.
Our attack was unable to differentiate between the \texttt{read-only} and \texttt{read-exec}, but it did detect additional pages (\texttt{0x55892ba96000} and \texttt{0x7f3eef13b000}) that had never been identified with a \texttt{/proc/PID/maps} file.
We investigated page tables using a custom kernel module and confirmed that all the detected permissions are correct.
The average runtime is 95 $sec.$ (51 $sec.$ for the masked load and 44 $sec.$ for the store).
The runtime can significantly be improved in the desktop processor.

\subsection{Attacks on Windows 10}
\label{subsec:majoros}

In Windows 10, the kernel and drivers are located between \texttt{0xfffff8}\textbf{\texttt{00000}}\texttt{00000}-\texttt{0xfffff8}\textbf{\texttt{80000}}\texttt{00000} with a 2 MiB boundary, which leads to 262144 possible offsets (i.e., 18 bits of entropy).
The entry point of the kernel is randomized within this address range and can begin at any 4-KiB boundary.
With this, we probed the kernel address space on an Intel i5-12400F.
As a result, we found the kernel address region--which is allocated in five consecutive 2-MiB pages--within 60 $ms$, on average.
In our attack, we only find the base address of the large region containing the kernel image.
However, this still derandomizes 18 bits of KASLR entropy and can be used in combination with our TLB attack (\textbf{$\varmathbb{P}4$}) to break the remaining 9 bits of entropy.
Additionally, we further conducted our attack on the KVAS-enabled Windows.
In Windows 10 (ver. 1709), the KVAS code (e.g., \texttt{KiSystemCall64Shadow}) is part of the kernel (as in Linux), and the offset from the kernel base address is \texttt{0x298000}.
On an Intel i7-6600U (Skylake), we probed the kernel address space with a 4-KiB alignment.
Consequently, we were able to find the KVAS region consisting of three consecutive 4-KiB pages in 8 $sec.$ with 100\% accuracy.
Thereafter, we found the kernel base address by subtracting the KVAS offset from the identified address.

\subsection{Breaking KASLR in cloud computing systems}
\label{subsec:kaslrcloud}

We conducted the attack on global cloud services, Amazon EC2 (Xeon E5-2676), Google GCE (Xeon Cascade Lake), and Microsoft Azure (Xeon Platinum 8171M).
Like with KASLR breaks before, we probed all possible kernel offsets in the range of randomization.
In all experiments, we successfully identified the kernel's base address as well as the currently loaded modules.
On Amazon EC2 (Linux kernel 5.11.0-1020-aws), since the processor is vulnerable to Meltdown, we found the KPTI trampoline at offset \texttt{0xe00000}, from which we were able to calculate the kernel's base address.
The runtime is 0.03 $ms$ for identifying the kernel base and 1.14 $ms$ for the kernel modules.
On Google GCE (Linux kernel  5.13.0), we identified the kernel base in 0.08 $ms$ and the kernel modules in 2.7 $ms$.
On Microsoft Azure (Windows 10, 21H2), we derandomized 18 bits of KASLR entropy in 2.06 $sec.$ (\textit{n = 1000}).

%% file: disc.tex
\section{Countermeasures}
\label{sec:disc}


\subsection{Software-based mitigations}
With Function Granular KASLR (FGKASLR)~\cite{fgkaslr}, individual kernel functions are reordered so that even if the kernel address is revealed, the attacker cannot identify the location of specific functions based on relative addresses.
However, even with FGKASLR enabled, our attack can still break the fine-grained KASLR by leveraging TLB state template attacks as described in~\cite{lipp2022amd}.
We further verified whether our attack could bypass FLARE~\cite{canella2020kaslr}, which maps dummy physical pages to mitigate KASLR breaks based on page mappings.
On the FLARE-enabled kernel, we mounted TLB (\textbf{$\varmathbb{P}4$}) and page-table (\textbf{$\varmathbb{P}2$}) attacks.
As a result, we could clearly identify the fast access times that reveal the mapped kernel regions in the TLB attack. 
While dummy mappings mitigate attacks based on the page table level, they do not prevent our TLB-based KASLR break.
Stronger isolation or re-randomization~\cite{williams2016shuffler} should be implemented to mitigate our attack successfully.
\vspace{0.5em}

\subsection{Hardware-based mitigations}
Since our TLB attack is based on TLBs, splitting TLB sets for user and kernel space can be used to mitigate our attack.
However, this mitigation is not practically possible since the partitioned TLBs do not fully support continuous virtual addresses, and it requires expensive hardware changes~\cite{gras2018translation,koschel2020tagbleed}.
Additionally, it is possible to replace the masked load and store instructions with \texttt{NOP}s only when the mask bits are all set to zero.
In Ubuntu 20.04.3 (kernel 5.11.0-27) with the default installation, we found only 6 out of 4104 executables that contain the masked load or store instruction, thus we believe that the solution of restricting or replacing masked operations has little impact on the system. 
We leave the detailed performance evaluation of these mitigations for future work.

%% file: related.tex
\section{Related works}
\label{sec:related}

\subsection{Microarchitectural attacks on KASLR}

Hund et al.~\cite{hund2013practical} introduced the first microarchitectural KASLR break by exploiting TLB states.
Jang et al.~\cite{jang2016breaking} significantly improved Hund et al.'s attack~\cite{hund2013practical} by using an Intel TSX.
Kosched et al.~\cite{koschel2020tagbleed} exploited tagged TLBs and data caches to break KASLR even in the KPTI-enabled kernel.
Gruss et al.~\cite{gruss2016prefetch} and Lipp et al.~\cite{lipp2022amd} exploited software prefetch, and Schwarzl et al.~\cite{schwarzl2021speculative} revisited the prefetch attack~\cite{gruss2016prefetch}.
Evtyushkin et al.~\cite{evtyushkin2016jump} leveraged collisions within BTB and Schwarz et al.~\cite{schwarz2019store} and Canella et al.~\cite{canella2019fallout} exploited a store-to-load forwarding optimization.
Lipp et al.~\cite{lipp2021platypus} introduced the first attack that solely uses power consumption differences.
Lipp et al.~\cite{lipp2022amd} exploited power variations of the prefetch instructions on AMD CPUs.
Canella et al.~\cite{canella2020kaslr} exploited an incomplete hardware fix for Meltdown and Weber et al.~\cite{weber2021osiris} exploited a cache line conflict caused by the non-temporal moves (\texttt{MOVNT}).

\subsection{AVX side-channel attacks}

Gruss et al.~\cite{schwarz2019netspectre} introduced the first AVX-based covert-channel, which is based on the timing differences in AVX2 power saving feature.
Ragab et al.~\cite{ragab2021rage} discovered that \texttt{VMASKMOV} instructions (with all-zero masks) that access invalid addresses issue a machine clear.
Weber et al.~\cite{weber2021osiris} discovered a timing side-channel that consists of \texttt{VDMADD132PD} and \texttt{FISTP} instructions as part of the result of their side-channel fuzzing framework.
In this paper, our side-channel attack exploits the fault-resistance and timing difference properties of the AVX masked operations.

%% file: concl.tex
\section{Conclusion}
\label{sec:concl}

This paper introduced a novel AVX timing side-channel attack that can defeat User or Kernel ASLR.
We demonstrated User and Kernel ASLR breaks on popular OSes, cloud computing systems, and an SGX enclave.
We also showed that our attack can effectively infer user behavior.
Our attack is very fast, reliable, and works on the vast majority of modern processors.
We highlight that stronger isolation or re-randomization should be implemented to successfully mitigate our presented attack.

%% file: main.bbl
\begin{thebibliography}{10}
\providecommand{\url}[1]{#1}
\csname url@samestyle\endcsname
\providecommand{\newblock}{\relax}
\providecommand{\bibinfo}[2]{#2}
\providecommand{\BIBentrySTDinterwordspacing}{\spaceskip=0pt\relax}
\providecommand{\BIBentryALTinterwordstretchfactor}{4}
\providecommand{\BIBentryALTinterwordspacing}{\spaceskip=\fontdimen2\font plus
\BIBentryALTinterwordstretchfactor\fontdimen3\font minus
  \fontdimen4\font\relax}
\providecommand{\BIBforeignlanguage}[2]{{%
\expandafter\ifx\csname l@#1\endcsname\relax
\typeout{** WARNING: IEEEtranS.bst: No hyphenation pattern has been}%
\typeout{** loaded for the language `#1'. Using the pattern for}%
\typeout{** the default language instead.}%
\else
\language=\csname l@#1\endcsname
\fi
#2}}
\providecommand{\BIBdecl}{\relax}
\BIBdecl

\bibitem{fgkaslr}
K.~C. Accardi, ``{Function Granular KASLR},''
  \url{https://lwn.net/Articles/824307/}, 2020.

\bibitem{mediainst2021amd}
AMD, ``{AMD64 Architecture Programmer's Manual Volume 4: 128-Bit and 256-Bit
  Media Instructions},'' \emph{AMD64 Technology}, 2021.

\bibitem{barr2010translation}
T.~W. Barr, A.~L. Cox, and S.~Rixner, ``{Translation Caching: Skip, Don't Walk
  (the Page Table)},'' \emph{ACM SIGARCH Computer Architecture News}, vol.~38,
  no.~3, pp. 48--59, 2010.

\bibitem{canella2019fallout}
C.~Canella, D.~Genkin, L.~Giner, D.~Gruss, M.~Lipp, M.~Minkin, D.~Moghimi,
  F.~Piessens, M.~Schwarz, B.~Sunar \emph{et~al.}, ``{Fallout: Leaking Data on
  Meltdown-resistant CPUs},'' in \emph{26th ACM Conference on Computer and
  Communications Security (CCS)}, 2019.

\bibitem{canella2020kaslr}
C.~Canella, M.~Schwarz, M.~Haubenwallner, M.~Schwarzl, and D.~Gruss, ``{KASLR:
  Break It, Fix It, Repeat},'' in \emph{15th ACM Asia Conference on Computer
  and Communications Security (ASIACCS)}, 2020.

\bibitem{kpti2017lwn}
J.~Corbet, ``{The current state of kernel page-table isolation},''
  \url{https://lwn.net/Articles/741878/}, 2017.

\bibitem{deilmann2012guide}
M.~Deilmann \emph{et~al.}, ``{A Guide to Vectorization with Intel® C++
  Compilers},'' \emph{Intel Corporation}, pp. 20--21, 2012.

\bibitem{evtyushkin2016jump}
D.~Evtyushkin, D.~Ponomarev, and N.~Abu-Ghazaleh, ``{Jump over ASLR: Attacking
  branch predictors to bypass ASLR},'' in \emph{49th Annual IEEE/ACM
  International Symposium on Microarchitecture (MICRO)}, 2016.

\bibitem{gras2018translation}
B.~Gras, K.~Razavi, H.~Bos, C.~Giuffrida \emph{et~al.}, ``{Translation
  Leak-aside Buffer: Defeating Cache Side-channel Protections with TLB
  Attacks},'' in \emph{27th USENIX Security Symposium (USENIX Security)}, 2018.

\bibitem{gruss2017kaslr}
D.~Gruss, M.~Lipp, M.~Schwarz, R.~Fellner, C.~Maurice, and S.~Mangard, ``{KASLR
  is Dead: Long Live KASLR},'' in \emph{9th International Symposium on
  Engineering Secure Software and Systems (ESSoS)}, 2017.

\bibitem{gruss2016prefetch}
D.~Gruss, C.~Maurice, A.~Fogh, M.~Lipp, and S.~Mangard, ``{Prefetch
  Side-Channel Attacks: Bypassing SMAP and Kernel ASLR},'' in \emph{23rd ACM
  Conference on Computer and Communications Security (CCS)}, 2016.

\bibitem{hund2013practical}
R.~Hund, C.~Willems, and T.~Holz, ``{Practical Timing Side Channel Attacks
  Against Kernel Space ASLR},'' in \emph{34th IEEE Symposium on Security and
  Privacy (S\&P)}, 2013.

\bibitem{intelsdmvol3a}
Intel, ``{Intel® 64 and IA-32 Architectures Software Developer’s Manual
  Volume 3A: System Programming Guide, Part 1},'' 2016.

\bibitem{intel2021optz}
------, ``{Intel® 64 and IA-32 Architectures Optimization Reference Manual},''
  2021.

\bibitem{intel2018mitigations}
------, ``{Speculative Execution Side Channel Mitigations},'' 2021.

\bibitem{appledoublemap}
A.~Ionescu, ``{Apple Double Map},''
  \url{https://twitter.com/aionescu/status/948609809540046849}, 2018.

\bibitem{jang2016breaking}
Y.~Jang, S.~Lee, and T.~Kim, ``{Breaking Kernel Address Space Layout
  Randomization with Intel TSX},'' in \emph{23rd ACM Conference on Computer and
  Communications Security (CCS)}, 2016.

\bibitem{koschel2020tagbleed}
J.~Koschel, C.~Giuffrida, H.~Bos, and K.~Razavi, ``{TagBleed: Breaking KASLR on
  the Isolated Kernel Address Space using Tagged TLBs},'' in \emph{IEEE
  European Symposium on Security and Privacy (EuroS\&P)}, 2020.

\bibitem{strictkernelmemory}
Z.~Li, ``{Support strict kernel memory permissions for security},''
  \url{https://lwn.net/Articles/812633/}, 2020.

\bibitem{lipp2022amd}
M.~Lipp, D.~Gruss, and M.~Schwarz, ``{AMD Prefetch Attacks through Power and
  Time},'' in \emph{31st USENIX Security Symposium (USENIX Security)}, 2022.

\bibitem{lipp2021platypus}
M.~Lipp, A.~Kogler, D.~Oswald, M.~Schwarz, C.~Easdon, C.~Canella, and D.~Gruss,
  ``{PLATYPUS: Software-based Power Side-Channel Attacks on x86},'' in
  \emph{IEEE Symposium on Security and Privacy (S\&P)}, 2021.

\bibitem{windowskvashadow}
{Microsoft Security Response Center}, ``{KVA Shadow: Mitigating Meltdown on
  Windows},'' 2018.

\bibitem{ragab2021rage}
H.~Ragab, E.~Barberis, H.~Bos, and C.~Giuffrida, ``{Rage Against the Machine
  Clear: A Systematic Analysis of Machine Clears and Their Implications for
  Transient Execution Attacks},'' in \emph{30th USENIX Security Symposium
  (USENIX Security)}, 2021.

\bibitem{schwarz2019store}
M.~Schwarz, C.~Canella, L.~Giner, and D.~Gruss, ``{Store-to-Leak Forwarding:
  Leaking Data on Meltdown-resistant CPUs (Updated and Extended Version)},''
  \emph{arXiv preprint arXiv:1905.05725}, 2019.

\bibitem{schwarz2019netspectre}
M.~Schwarz, M.~Schwarzl, M.~Lipp, J.~Masters, and D.~Gruss, ``{NetSpectre: Read
  Arbitrary Memory over Network},'' in \emph{24th European Symposium on
  Research in Computer Security (ESORICS)}, 2019.

\bibitem{schwarzl2021speculative}
M.~Schwarzl, T.~Schuster, M.~Schwarz, and D.~Gruss, ``{Speculative
  Dereferencing: Reviving Foreshadow},'' in \emph{25th International Conference
  on Financial Cryptography and Data Security (FC)}, 2021.

\bibitem{shacham2007geometry}
H.~Shacham, ``{The Geometry of Innocent Flesh on the Bone: Return-into-libc
  without Function Calls (on the x86)},'' in \emph{14th ACM Conference on
  Computer and Communications Security (CCS)}, 2007.

\bibitem{weber2021osiris}
D.~Weber, A.~Ibrahim, H.~Nemati, M.~Schwarz, and C.~Rossow, ``{Osiris:
  Automated Discovery of Microarchitectural Side Channels},'' in \emph{30th
  USENIX Security Symposium (USENIX Security)}, 2021.

\bibitem{williams2016shuffler}
D.~Williams-King, G.~Gobieski, K.~Williams-King, J.~P. Blake, X.~Yuan, P.~Colp,
  M.~Zheng, V.~P. Kemerlis, J.~Yang, and W.~Aiello, ``{Shuffler: Fast and
  Deployable Continuous Code Re-Randomization},'' in \emph{12th USENIX
  Symposium on Operating Systems Design and Implementation (OSDI)}, 2016.

\end{thebibliography}
